\documentclass[12pt]{iopart}

\usepackage{iopams}
\usepackage{amssymb}
\usepackage{amsopn}
\usepackage[english]{babel}
\usepackage[latin1]{inputenc}
\usepackage[T1]{fontenc}
\usepackage{xy}
\usepackage{graphicx}
\usepackage{verbatim}
\usepackage{enumerate}
\usepackage{url}
\usepackage{subfig}
\usepackage{bm}
\usepackage{stackrel}

\usepackage{txfonts}

\newcommand{\imagescaling}{0.95}

\newcommand{\eqref}[1]{\eref{#1}}
\newcommand{\dotsc}{\ldots}
\newcommand{\text}{\textrm}

\DeclareMathOperator{\mod}{mod}

\DeclareMathOperator*{\argmax}{arg\,max}

\newcommand{\maxdim}{N}

\newcommand{\Z}{\mathbb{Z}}

\newcommand{\R}{\mathbb{R}}

\newcommand{\Tor}{\mathbf{T}}

\renewcommand{\S}{\mathbf{S}}

\newcommand{\B}{\mathbf{B}}
\newcommand{\A}{\mathbf{A}}

\newcommand{\OP}{Z}
\newcommand{\OPh}{{\OP}\ind{d}}
\newcommand{\aOP}{R}
\newcommand{\aOPh}{\aOP\ind{d}}

\newcommand{\ud}{\mathrm{d}}

\newcommand{\sset}[1]{\left\lbrace #1\right\rbrace}

\newcommand{\abs}[1]{\left| #1 \right|}

\newcommand{\ind}[1]{_{\mathrm{#1}}}
\newcommand{\vphi}{\varphi}

\renewcommand{\a}{{a}}
\newcommand{\xt}{x_*}

\newcommand{\Ob}{Q}
\newcommand{\Obf}{q}

\newcommand{\V}{\nu}
\newcommand{\Dt}{\Delta t}

\renewcommand{\mathbb}{\varmathbb}

\begin{document}


\title{Controlling Chimeras}
\author{Christian Bick\footnote{Present address: College of Engineering, Mathematics, and Physical Sciences, University of Exeter, North Park Road, Exeter EX4 4QF, United Kingdom.}\textsuperscript{$a,b$} 
and Erik A.~Martens\textsuperscript{${c,d,e}$}%
}

\address{${}^a$Department of Mathematics, Rice University, MS--136, 6100 Main St, Houston, TX 77005, USA}
\address{${}^{b}$Network Dynamics, Max Planck Institute for Dynamics and Self-Organization, 37077 G\"ottingen, Germany}
\address{${}^c$Department of Biomedical Science, University of Copenhagen, Blegdamsvej 3, 2200 Copenhagen, Denmark}
\address{${}^d$Department of Mathematical Sciences, University of Copenhagen, Universitetsparken 5, 2200 Copenhagen, Denmark}
\address{${}^{e}$Biophysics and Evolutionary Dynamics, Max Planck Institute for Dynamics and Self-Organization, 37077 G\"ottingen, Germany}
\date{\today}



\begin{abstract}
Coupled phase oscillators model a variety of dynamical phenomena in nature and technological applications. Non-local coupling gives rise to chimera states which are characterized by a distinct part of phase-synchronized oscillators while the remaining ones move incoherently. Here, we apply the idea of control to chimera states: using gradient dynamics to exploit drift of a chimera, it will attain any desired target position. Through control, chimera states become functionally relevant; for example, the controlled position of localized synchrony may encode information and perform computations. Since functional aspects are crucial in (neuro-)biology and technology, the localized synchronization of a chimera state becomes accessible to develop novel applications. Based on gradient dynamics, our control strategy applies to any suitable observable and can be generalized to arbitrary dimensions. Thus, the applicability of chimera control goes beyond chimera states in non-locally coupled systems.
\end{abstract}

\pacs{05.45.-a, 05.45.Gg, 05.45.Xt, 02.30.Yy}
\maketitle


\section{Introduction}
Collective behavior emerges in a broad range of oscillatory
systems in nature and technological applications. Examples include 
flashing fireflies, superconducting Josephson junctions, oscillations 
in neural circuits and chemical reactions, and many 
others~\cite{Pikovsky2003, Strogatz2004}. Phase coupled oscillators
serve as paradigmatic models to study the dynamics of such 
systems~\cite{Kuramoto, Acebron2005, Strogatz2000, Strogatz2001}. 
Remarkably, localized synchronization---in contrast to global 
synchrony---may arise in non-locally coupled systems where the coupling 
depends on the spatial distance between two oscillators. Dynamical states
consisting of locally phase-coherent and incoherent parts have been 
referred to as 
chimera states~\cite{Kuramoto2002, Abrams2004}, alluding to the 
fire-breathing Greek mythological creature composed of incongruous 
parts from different animals. Chimera states are relevant in a range
of systems; they have been observed experimentally in 
mechanical, \mbox{(electro-)}chemical, and laser
systems~\cite{Martens2013, Wickramasinghe2013, Tinsley2012, Hagerstrom2012}, 
and related localized activity has been associated with neural 
dynamics~\cite{Amari1977, Zhang1996, Compte2000, Laing2001, Laing2009b, Sakaguchi2006a, Olmi2010, Omelchenko2013, Hizanidis2013, Wildie2012, Tognoli2014, Pazo2014}.
By definition, local synchrony is tied to a spatial position that 
may directly relate to function: in a neural network, for example, 
different neurons encode different 
information~\cite{Hubel1959, Fyhn2004, Martens2010bistable}. In non-locally 
coupled phase oscillator rings, the spatial position of partial 
synchrony not only depends strongly on the initial conditions~\cite{Kuramoto2002}, 
but it also is subject to pseudo-random (i.e., low number) 
fluctuations~\cite{Omel'chenko2010}. These fluctuations are particularly 
strong for persistent chimeras for just few oscillators~\cite{Sieber2014} 
as in typical experimental setups. This naturally leads to the question 
whether it is possible to control a chimera state and keep at a desired 
spatial location.

In this article, we derive a control scheme to dynamically modulate 
the position of the coherent part of a chimera. To the best of our 
knowledge, this is the first application of noninvasive control to spatial 
properties of chimera states. Our control is based on gradient dynamics to
optimize general location-dependent averages of dynamical 
states. Defined as the place where local synchronization is maximal, the 
spatial location of a chimera state is such a space-dependent average. As 
control of spatially localized patterns, chimera control relates---by 
definition---to both traditional control 
approaches~\cite{Fradkov1998, Insperger2013, Ott1990} 
as well as other of localized patterns in, for example, 
chemical~\cite{Mikhailov2006} or optical~\cite{Ayoub2011} systems.
However, the aim of chimera control differs from these control 
approaches. First, chimera control preserves a chimera state as a 
whole as opposed to classical engineering control. More specifically, 
its goal is not to change the dynamics qualitatively, that is, for 
example, to restore a turbulent system to a periodic state, but rather 
to control space dependent averages. Second, chimera control is 
noninvasive as a result of the underlying gradient dynamics. That is,
in contrast to some approaches to control the spatial position of 
localized patterns~\cite{Ayoub2011}, the control strength vanishes upon 
convergence. Third, chimera control extends beyond the spatial 
continuum limit, where the dynamics of individual oscillators are 
negligible. It applies to systems of finite dimension, even down to just 
a handful of inhomogeneous oscillators. In contrast to continuous spatial 
systems, where static or periodic localized 
patterns~\cite{Burke2007, Umbanhowar1996, Zhang1996, Lober2014} may 
shift, chimeras in finite dimensions are localized chaotic 
states~\cite{Wolfrum2011a} (similar 
to localized turbulence in pipe flows~\cite{Prigent2002, Barkley2005}) 
subject to strong low number fluctuations~\cite{Omel'chenko2010}. In 
summary, chimera control 
modulates the spatial location of a chimera noninvasively even in 
low-dimensional system and preserves its ``internal'' incoherent 
oscillatory dynamics.

We anticipate chimera control to have a broad impact across 
different fields. On the one hand, the control scheme may elucidate 
how position is maintained in (noisy and heterogeneous) real world 
systems where spatial localization of synchrony plays a functional 
role, such as neural systems. On the other hand, however, it is 
the first step towards actually employing chimera states as 
functional localized spatio-temporal patterns. In fact, instead of 
passively observing chimera states, the aim of control is to actively 
exploit chimeras 
for applications by making the spatial location accessible. Its 
location could encode information which allows, for example, control 
mediated computation. Despite the differences to chimera control, the
control of dynamical states, such as chaos, has led to many intriguing 
applications in their own right~\cite{Ott1990, Garfinkel1992, Scholl2007, Steingrube2010}. 
So in analogy to the Greek mythological creature one may ask: what 
would you be able to do if you could control a fire-breathing chimera?


\section{Chimeras in Non-Locally Coupled Rings}%
Rings of non-locally coupled phase oscillators provide a well studied
model in which chimera states may occur~\cite{Abrams2004}.
Let $\S:=\R/\Z$ be the unit interval with endpoints identified
and let $\Tor:=\R/2\pi\Z$ denote the unit circle. Let~$d$ be a 
distance function on~$\S$, $h:\R\to\R$ be a positive function, and 
$\alpha\in\Tor$, $\omega\in\R$ be parameters. The dynamics
of the oscillator at position $x\in\S$ on the ring is given by
\begin{equation}\label{eq:ContinuumLimit}
\partial_t\vphi(x,t) = \omega - \int_{0}^1 h(d(x,y)) \sin\!\left(\vphi(x, t)-\vphi(y, t)+\alpha\right)\ud y.
\end{equation}
The coupling kernel~$h$ determines the interaction strength 
between two oscillators depending on their mutual distance. 
The system evolves on the torus $\S\times\Tor$ where $x\in\S$
is the spatial position of an oscillator on the ring and 
$\vphi(x, t)\in\Tor$ its phase at time~$t$ on the torus. 

\begin{figure}
\begin{center}\includegraphics[scale=\imagescaling]{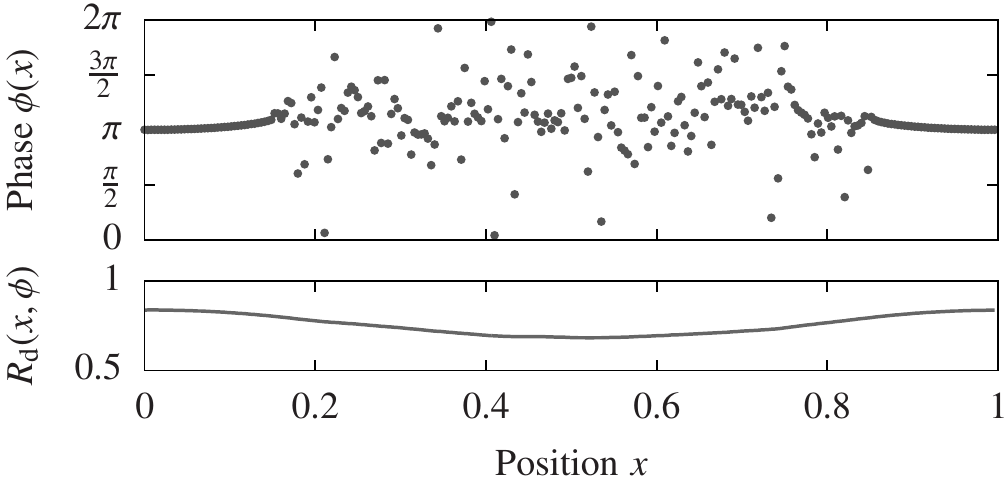}\end{center}
\caption{\label{fig:Chimera}
The local order parameter $\aOP(x, \phi)$ encodes the spatial position 
of a chimera state in a ring of $\maxdim=256$ oscillators. Non-local 
coupling is given by the exponential kernel~$h_0$; 
cf.~\eqref{eq:CouplingKernel}. As a function of the oscillator phase~$\phi(x)$ 
on the circle~$\S$ (top panel), the maximum of~$\aOP$ indicates the center 
of the synchronized region, the minimum the position of the incoherent part 
(bottom panel).
}
\end{figure}

Chimera states are characterized by a region of local phase 
coherence while the rest of the oscillators rotate incoherently.
Let $\phi\in\Phi := \sset{\phi: \S\to\Tor}$ denote a configuration
of phases on the ring. The local order parameter
\begin{equation}\label{eq:LocalOrderParam}
\OP(x, \phi) = \int_{0}^1 h(d(x, y)) \exp(i\phi(y))\ud y
\end{equation}
is an observable which encodes the local level of synchrony of~$\phi$
at $x\in\S$. That is, its absolute 
value $\aOP(x, \phi) = \abs{\OP(x, \phi)}$ is close to
zero if the oscillators are locally spread out and 
attains its maximum if the phases are phase synchronized close 
to~$x$. A chimera state is a solution~$\vphi(x, t)$ of~\eqref{eq:ContinuumLimit}
which consists of locally synchronized and locally incoherent parts. 
The value of the local order parameter yields local properties of a 
chimera. The local order parameter obtains its maximum at the center of the
phase synchronized region and its minimum at the center of the incoherent 
region; cf.~Figure~\ref{fig:Chimera} for a finite dimensional 
approximation. 

\section{Chimera Control}%

Is it possible to dynamically move a state to a desired 
position by exploiting drift properties? Before considering chimera
states, we consider general solutions moving in space. Here we
focus on systems with one spatial dimension but it is straightforward 
to extend the notions to higher dimensions. A solution of~\eqref{eq:ContinuumLimit}
may be seen as a one-parameter family of functions~$\vphi_t\in\Phi$ 
which assign a phase to each spatial position. Let $\Ob:\S\times\Phi\to\R^n$ 
be differentiable in the first argument. Think of~$\Ob$ as an observable 
of the system which depends on the spatial position~$\S$; here we look 
at the particular circular geometry because of its relevance in the 
context of chimera states on a ring, but one could as well consider 
observables on other geometries, such as the line~$\R$. A solution~$\vphi_t$ 
of~\eqref{eq:ContinuumLimit} with initial condition $\vphi_0\in\Phi$
is called $\Ob$-traveling along~$\S$ if there are suitably smooth
functions~$y(t)$ and $\Obf:\S\to\R^n$ such that 
$\Ob(x, \vphi_t) = \Obf(x-y(t))$ for all~$t$; in particular, a solution is 
$\Ob$-traveling at constant speed $v\in\R$ along~$\S$ if
$\Ob(x, \vphi_t) = \Obf(x-vt)$ for all~$t$. Hence, the temporal 
evolution of a $\Ob$-traveling solutions in terms of the
observable~$\Ob$ is a shift along~$\S$.

If there is a way to influence the spatial motion in a controlled way, it 
can be used to optimize a general observable~$\Ob$. Let $\partial_zf(z)|_{z_0}$ 
denote the partial derivative of a function~$f$ with respect to~$z$ 
at~$z_0$, let $f'$ denote its total derivative, and~$\dot z$ the 
temporal derivative of a function~$z(t)$. Let~$\vphi_t$ be a $\Ob$-traveling 
solution with~$\Obf(x)$ and~$y(t)$ such that $\Ob(x, \vphi_t) = \Obf(x-y(t))$. 
The function~$y(t)$ describes the spatial position of~$\vphi_t$ with 
respect to~$\Ob$. For now, fix a target $\xt\in\S$ and assume that~$\Obf$ 
is differentiable with all critical points being extrema. The idea is to 
use an accessible system parameter that governs the evolution 
of~$\vphi_t$ in terms of the observable~$\Ob$ to maximize~$\Ob$ 
at~$\xt$, or, put differently, to use the knowledge how this accessible 
system parameter influences the evolution of~$y(t)$ to maximize 
$\Obf_{\xt}(y):=\Obf(\xt - y)$ in~$y$. To this end, we assume that for 
a given observable~$\Ob$ there is a family~$h_\a$ of coupling kernels, 
indexed by $\a\in A\subset\R$, and a continuous invertible map 
$\V:A\to\R$ such that~$\vphi_t$ is a $\Ob$-traveling 
solution at speed~$\dot y=\V(\a)$ of~\eqref{eq:ContinuumLimit} with 
coupling kernel~$h_a$. In other words, we assume that the position~$y(t)$ 
of the solution~$\vphi_t$ is given by integrating~$\V(\a)$. Of course, 
if~$\a$ is constant we have $\Ob(x,\vphi_t)=\Obf(x-\V(\a)t)$, i.e., 
$\vphi_t$ is $\Ob$-traveling at constant speed along~$\S$.

Control can now be realized as gradient dynamics by choosing the 
parameter~$\a$ suitably. For $\gamma>0$ and assuming that the initial 
condition is not a local minimum, the function~$\Obf_{\xt}(y)$ is 
maximized if~$y$ is subject to 
the gradient dynamics ${\dot y} = \gamma\partial_y\Obf_{\xt}(y)$ since this 
choice implies that $\dot\Obf_{\xt}\geq 0$. Note that 
$\partial_y\Obf_{\xt}(y)=-\Obf'(\xt-y)=-\partial_x\Obf(x-y)|_{\xt}$. 
Thus, if the function~$y(t)$ of a $\Ob$-traveling solution~$\vphi_t$ obeys
\begin{equation}\label{eq:GradientDynamics2}
{\dot y} = -\gamma\partial_x\Obf(x-y)|_{\xt}=-\gamma\partial_x\Ob(x, \vphi_t)|_{\xt},
\end{equation}
then the function $\Ob(x, \vphi_t)$ will attain a (local) maximum 
at~$x=\xt$ in the limit of $t\to\infty$. At the same time, the 
map~$\V$ allows to 
use~$\a$ as a control parameter. By definition we have
$\dot y(t)=\V(a(t))$ and therefore~\eqref{eq:GradientDynamics2} yields
\begin{equation}\label{eq:ControlDynamicsGen}
a(t) = \V^{-1}\left(-\gamma\partial_x\Ob(x, \vphi_t)|_{\xt}\right),
\end{equation}
a direct relationship between the traveling solution and the 
parameter~$\a$. More precisely, choosing a time dependent control 
parameter~$\a$ according to~\eqref{eq:ControlDynamicsGen} yields
a traveling solution whose dynamics maximize the observable~$\Ob$ 
at~$\xt$.

Note that convergence to the target through control does not depend 
on the function~$\V$. Moreover, the assumption that~$\V$ is invertible
can be relaxed. If~$\V:A\to U$ be invertible where~$U\subset\R$ is an 
open interval that contains zero, then we can just extend~$\V^{-1}$ 
from~$U$ onto the real line~$\R$ by choosing 
$\V^{-1}(u)=\sup_{a\in A}\V^{-1}(a)$ for 
$u\geq\sup U$ and $\V^{-1}(u)=\inf_{a\in A}\V^{-1}(a)$ for $u\leq\inf U$ 
or vice versa. Effectively this yields gradient dynamics
${\dot y} = \gamma(t)\partial_y\Obf_{\xt}(y)$
with time-dependent parameter 
$0<\gamma(t)\leq\gamma$ which maximize~$\Obf_{\xt}$. Thus, with 
the assumptions on~$\V$ as above, control remains applicable. On the 
other hand, to determine the maximal convergence speed one has to to 
take other properties of~$\V$ into account.

The same gradient approach can be used to apply control to sufficiently
smooth time-dependent control targets. Even though we have so far 
assumed~$\xt$ to be constant, the control target can also be taken
to be piecewise constant since the values at the discrete points of
discontinuity do not change the integral. Therefore, control is suitable 
for any time dependent control target~$\xt(t)$ that can be approximated
by a piecewise constant functions. Of course, convergence to a time-dependent
control target will only be approximate as control ensures that 
the maximum is attained only in the limit as $t\to\infty$.

To control chimeras we apply this general control scheme to the
absolute value~$R$ of the local order parameter. Since it encodes the 
local level of synchrony, dynamics that maximize the local order 
parameter through $\aOP$-traveling chimera solutions yield a 
chimera moving to a specified target position. Note that 
$R(x, \vphi_t)=r(x)$ of a chimera state $\vphi_t$ is 
stationary~\cite{Abrams2004, Omel'chenko2013} so it is $R$-traveling 
at constant speed zero. Here we further assume that there is a family 
of coupling kernels~$h_\a$ that lead to $\aOP$-traveling solutions at 
nonzero speed $\V(\a)$. The control parameter 
dynamics~\eqref{eq:ControlDynamicsGen} for the observable~$\aOP$ are
\begin{equation}\label{eq:ControlDynamics}
a(t) = \V^{-1}\left(-\gamma\partial_x\aOP(x, \vphi_t)|_{\xt}\right).
\end{equation}
Hence, choosing a time dependent control parameter~$\a$ according 
to~\eqref{eq:ControlDynamics} is equivalent to gradient dynamics 
to maximize the local order parameter at~$\xt$. For the original
chimeras with a single coherent region~\cite{Kuramoto2002, Abrams2004},
i.e., where~$R$ has a global maximum, the limiting position of a 
chimera subject to control is unique. For chimera states with multiple 
coherent regions~\cite{Omelchenko2013, Omel'chenko2013}, the local 
order parameter will attain a local maximum at the target position.


\begin{figure*}
\begin{center}\includegraphics[scale=\imagescaling]{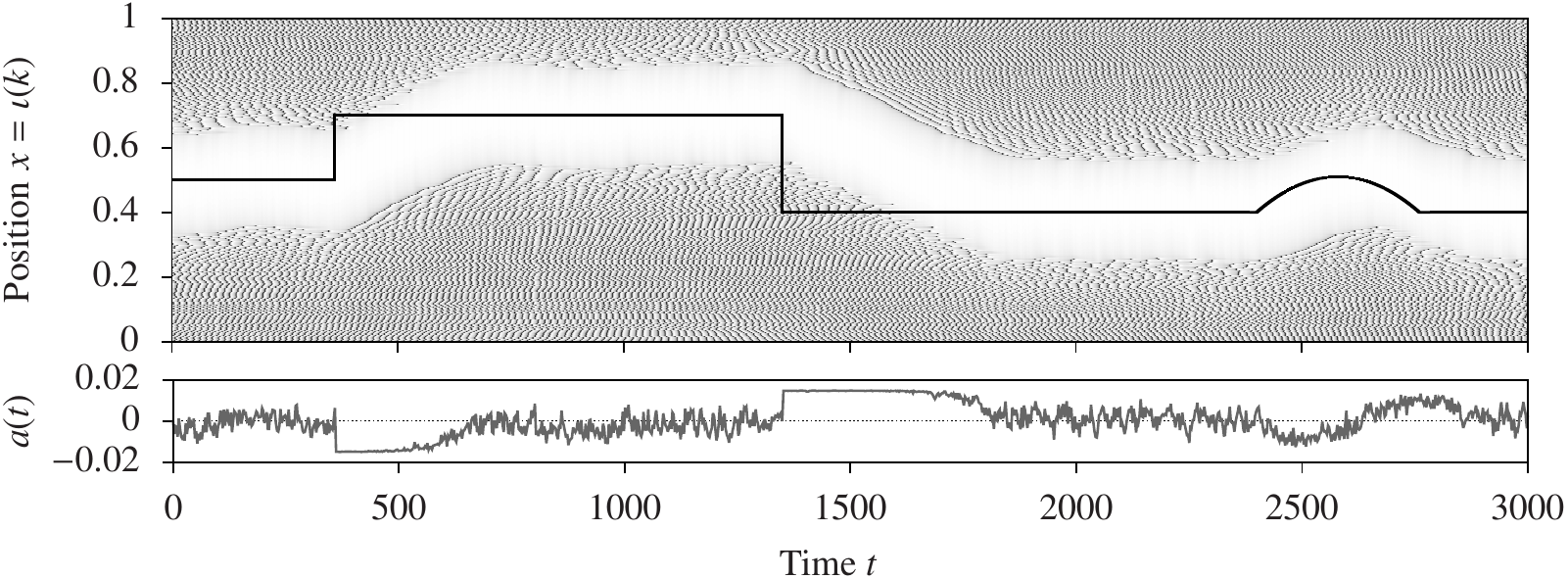}\end{center}
\caption{\label{fig:ControlledChimera}
The position of the chimera adjusts to the imposed target for the control 
scheme applied to a ring of $\maxdim=256$ oscillators. The top panels shows 
the phase evolution in the co-rotating frame defined by the phase in the 
synchronized region with maximal order parameter. The black line is the 
target position. The bottom panels depict the asymmetry parameter~$\a(t)$
bounded by $\a\ind{max}=0.015$, cf.~\eqref{eq:DiscrControlScheme}. Once the
target position is reached,~$\a$~stays close to zero.
}
\end{figure*}

\section{Implementation in Finite Dimensional Rings}%
\label{sec:Implementation}
Most real world systems consist of a finite number of oscillators; we thus
implement chimera control in an approximation of the continuous 
equations~\eqref{eq:ContinuumLimit} by a system of~$\maxdim$ phase 
oscillators. Let $\iota(k)=k/\maxdim$ be the position of the $k$th 
oscillator on the ring~$\S$. Let $\omega_k\in\R$ be the 
intrinsic frequency of each oscillator and initially we assume 
that the oscillator system is homogeneous, i.e., $\omega_k=\omega$ 
for all $k=1, \dotsc, \maxdim$. 
The temporal evolution of each oscillator is given by
\begin{equation}\label{eq:RingDynamics}
\dot\vphi_k = \omega_k - \frac{1}{\maxdim}\sum_{j=1}^\maxdim
h(d(\iota(k), \iota(j)))\sin(\vphi_k-\vphi_j+\alpha)
\end{equation}
for $k=1, \dotsc, \maxdim$. Here, 
$d(x, y) = \big(\big(x-y+\frac{1}{2}\big) \mod 1\big)-\frac{1}{2}$
is a signed distance function on~$\S$. The local order parameter 
of the discretized system is defined for 
$\vphi = (\vphi_1, \dotsc, \vphi_N)\in\Tor^\maxdim$ as
\begin{equation}\label{eq:LocalOrderParam2}
\OPh(x, \vphi) = \frac{1}{\maxdim}\sum_{j=1}^\maxdim h(d(x, \iota(j))) \exp(i\vphi_j)
\end{equation}
and its absolute value~$\aOPh(x, \vphi)$ encodes the local level 
of synchrony; cf.~Figure~\ref{fig:Chimera}.

To implement the chimera control scheme~\eqref{eq:ControlDynamicsGen}, 
the assumption of a monotonic relationship~$\nu$ between a system 
parameter and the chimera's drift speed has to be satisfied. 
Asymmetric coupling kernels may induce drift in dynamical systems 
on a continuum such as standard pattern forming 
systems~\cite{Zhang1996, Burke2009, Siebert2014}. We employ 
the recent observation that breaking the symmetry of the coupling kernel 
slightly also results in the drift of the chimeras in finite-dimensional 
systems~\cite{Bick2014}. The result is a monotonic relationship~$\nu(a)$ 
between asymmetry and drift speed~\cite{Bick2014} independent of the 
system's dimension. Here we consider a family of exponential coupling 
kernels
\begin{equation}\label{eq:CouplingKernel}
h_\a(x) = \cases{\exp\left(-\kappa(1-\a)\abs{x}\right) &\text{if } $x < 0$\\
\exp\left(-\kappa(1+\a)\abs{x}\right) &\text{if } $x \geq 0$.}
\end{equation}
for $\a\in(-1,1)$, where~$\a$ determines the symmetry of the coupling 
kernel. The coupling in~\eqref{eq:CouplingKernel} can be analytically 
related to oscillators coupled in reactive-diffusive media~\cite{Shima2004} 
subject to convective concentration gradients of the coupling medium. For 
sufficiently small $\abs{a}\lessapprox 0.015$
the relationship~$\V$ between drift and asymmetry is approximately linear 
at $\a=0$ and the resulting drifting chimeras 
are in good approximation $R$-traveling with constant speed. 
We use this single observation for the implementation of chimera control. Note
the particular shape of~$h_\a$ is not crucial for control since other asymmetric 
coupling kernels also lead to drift. However, the topic of drifting chimera states 
in systems with asymmetric coupling kernels deserves a treatment in its own 
right, and we refer to a forthcoming article~\cite{Bick2014} for details.

The relationship between asymmetry parameter~$\a$ and the drift speed 
now allows for a straightforward implementation of the control 
scheme. The control rule~\eqref{eq:ControlDynamics} acts as feedback
control through the asymmetry parameter. If the chimera is off target,
the nonzero asymmetry yields a drift of the chimera towards the target according
to the derivative of the local order parameter at the target position. 
Once the target is approached, the control subsequently reduces the 
asymmetry and acts a corrective term keeping the chimera on target. 
For the finite ring, a discrete derivative at $\xt\in\S$ can be defined for a given 
$\delta\in(0, 0.5)$ by
\begin{equation}
\Delta^\delta_{\xt} \aOPh(x, \vphi(t)) = \frac{1}{2\delta}\left({\aOPh(\xt+\delta,\vphi(t))}-{\aOPh(\xt-\delta,\vphi(t))}\right).\
\end{equation}
For small~$\delta$ we have 
$\Delta^\delta_{\xt} \aOPh(x, \vphi(t))\approx\partial_x\aOPh(x, \vphi(t))|_{\xt}$. 
We employ the sigmoidal function $\lambda(x) = 2(1+\exp(-x))^{-1}-1$ to 
ensure an upper bound $\a\ind{max} > 0$ for the asymmetry parameter~$\a(t)$
to prevent chimeras from breaking down. Let $K>0$ be a constant. Given 
a target position~$\xt\in\S$, an approximation 
of~\eqref{eq:ControlDynamics} for control is 
\begin{equation}\label{eq:DiscrControlScheme}
\a(t) = a\ind{max}
\lambda\left(K\Delta^\delta_{\xt} \aOPh(x, \vphi(t))\right).
\end{equation}
where~$K$ can be determined from the gradient control parameter 
$\gamma=K\V'(0)$. These dynamics will maximize the local order 
parameter at~$\xt$. In other words, a chimera~$\vphi(t)$ will move 
along the ring until its synchronized part is centered at~$\xt$.

Solving the dynamical equations subject to control numerically
shows that the chimera adjusts to the imposed target position.
Figure~\ref{fig:ControlledChimera} shows a simulation for 
$\maxdim=256$ phase oscillators with $K=100$, and a time dependent
target position~$\xt(t)$. The simulation is carried out with initial 
conditions as in~\cite{Abrams2004} and an adaptive integration step
to meet standard error tolerances.
We discretized~\eqref{eq:DiscrControlScheme} in time by keeping 
the asymmetry parameter piecewise constant with an update every 
$\Dt=1$ time units. 
The chimera tracks the changes of the target position and adjusts to 
match new control targets.

Effectively, the control can be seen as a coupling of the dynamical 
equations to a function of the local order parameter. In contrast 
to systems with symmetric order parameter-dependent
interaction~\cite{Rosenblum2007, Bordyugov2010}, in chimera control
the order parameter induces a time-dependent 
asymmetry~\eqref{eq:ControlDynamics} to the nonlocal coupling to 
realize directed motion~\cite{Bick2014}.
As a result, the chimera drifts along a subspace defined by the 
symmetry of the uncontrolled system to achieve the target position.


\section{Control of Fluctuations}%

\begin{figure}
\begin{center}\includegraphics[scale=\imagescaling]{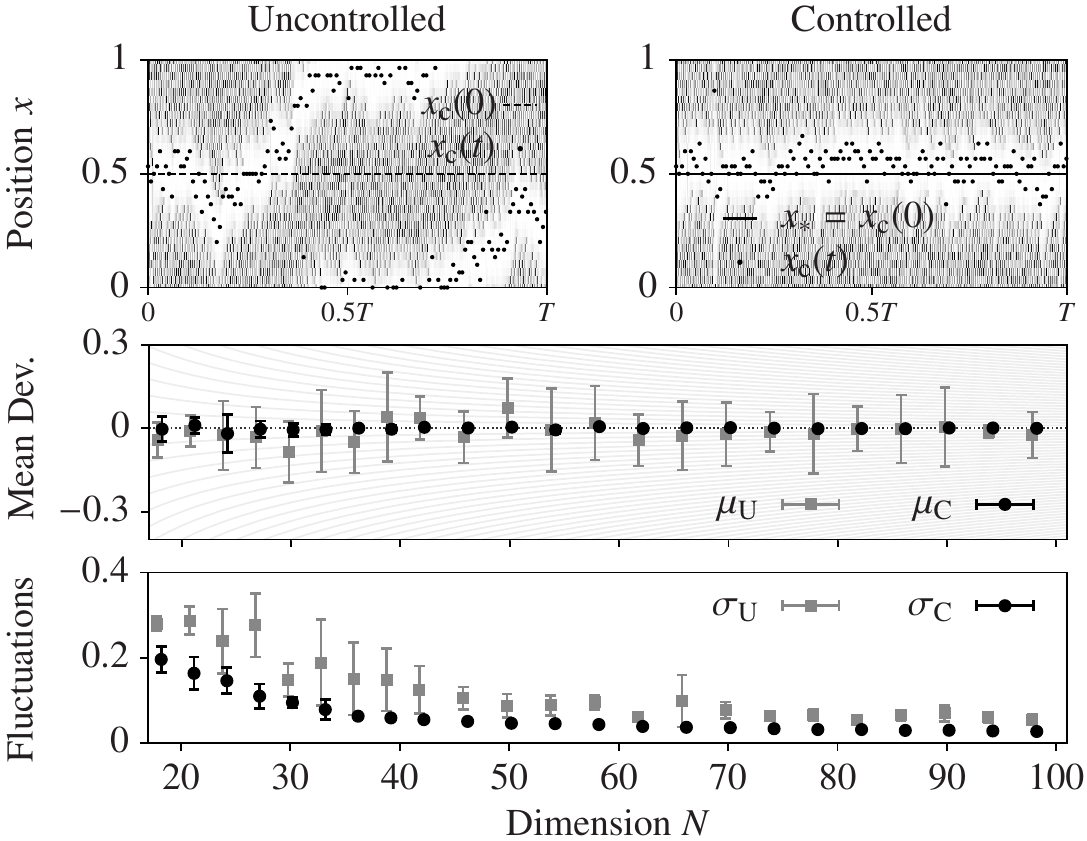}\end{center}
\caption{\label{fig:FluctuationControl}
Top: Control successfully suppresses pseudo-random finite size fluctuations
($\maxdim=30$ oscillators) in low-dimensional rings.  Center: The 
average deviation $\mu\ind{U}, \mu\ind{C}$ of a chimera from its initial 
position (straight lines in top panel) over $T=3000$ time units is distributed 
around zero without control (gray) but on spot with control (black). 
Thin gray lines indicate a deviation of a single oscillator.
Bottom: Control also reduces the fluctuations of the chimera due to pseudo-random 
movement significantly, even for very few oscillators ($N \leq 30$) as 
quantified by $\sigma\ind{U}, \sigma\ind{C}$. Points 
are slightly set off horizontally for legibility.}
\end{figure}

An uncontrolled chimera will exhibit pseudo-random  (low number)
fluctuations~\cite{Omel'chenko2010} along the ring~$\S$ which persist 
even when the symmetry of the system is broken. These fluctuations are 
particularly strong for small numbers of oscillators. Since chimera 
control acts as a feedback mechanism to correct deviations from the 
target position, it counteracts the fluctuations along the ring. Thus, 
the control scheme keeps a chimera localized at a target position even 
in low-dimension systems despite the strong spatial fluctuations for a
small number of oscillators; cf.~Figure~\ref{fig:FluctuationControl} 
(top).

To quantify how chimera control suppresses the pseudo-random 
fluctuations, we tracked the center of the coherent 
region~$x\ind{c}(t)\in\S$ in a 
homogeneous ring. More specifically, for a given initial condition 
$\vphi(0)$ for \eqref{eq:RingDynamics} with initial position $x\ind{c}(0)$
we first solved the uncontrolled system numerically to obtain the 
mean~$\mu\ind{U}$ and standard deviation $\sigma\ind{U}$ of 
$d(x\ind{c}(0), x\ind{c}(t))$ over~$T$ time units. Similarly, one 
obtains~$\mu\ind{C}$ and~$\sigma\ind{C}$ for the controlled chimera 
with $\xt=x\ind{c}(0)$ as the target position. Averages over multiple 
runs are shown in Figure~\ref{fig:FluctuationControl}. Applying 
control keeps the average position of the chimera on target for 
$\maxdim\geq 30$ (the standard deviation is below a single oscillator). 
Moreover, the fluctuations of the chimeras' positions are greatly 
reduced for all~$\maxdim$. Hence, control renders the spatial position
of a chimera usable even when the number of oscillators is small.

\section{Control for Inhomogeneous Rings}

For control to be relevant in real-world applications, it has to be
robust to inhomogeneities in the system. So far we have considered
the case of homogeneous rings where all oscillators have the same
intrinsic frequency~$\omega_k=\omega$ for $k=1, \dotsc, \maxdim$.
In fact, when all oscillators are identical, the ring has a 
rotational symmetry where the symmetry group acts by translations
along the ring. Control allows to shift a chimera along the orbit 
of the associated symmetry operation. Chimera states persist if the 
rotational symmetry is broken by choosing nonidentical 
frequencies, i.e. chimera solutions can be continued  while adiabatically increasing heterogeneity~\cite{Laing2009c}. Assume nonidentical intrinsic 
frequencies $\omega_k = 1+\eta_k$ where~$\eta_k$ are independently
sampled from a normal distribution centered at zero with standard 
deviation~$\sigma_\omega$. Chimeras can be observed for the 
inhomogeneous ring for 
$\sigma_\omega\lessapprox 0.03$ before the chimeras break down. In 
contrast to homogeneous oscillators, a chimera now has preferred 
positions on the inhomogeneous ring due to the lack of rotational 
symmetry which is determined by the actual value 
of the frequencies~$\omega_k$.

\begin{figure}
\begin{center}\includegraphics[scale=\imagescaling]{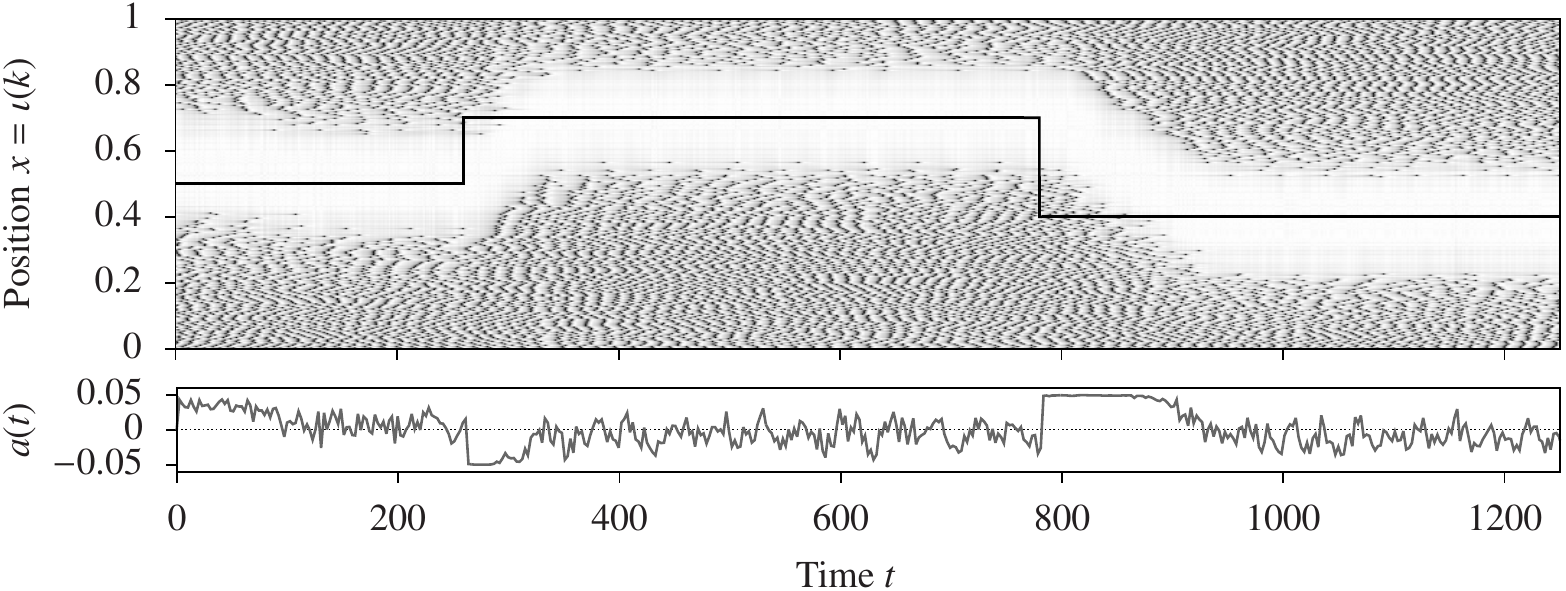}\end{center}
\caption{\label{fig:ControlledChimeraInhom}
Control of chimera states is successful even in heterogeneous rings of $\maxdim=256$ 
oscillators, yielding qualitatively similar results as for
homogeneous rings. As in Figure~\ref{fig:ControlledChimera} 
the phase of the oscillators in the top panel is shown in a 
corotating frame. The standard deviation of the oscillators' 
frequencies is~$\sigma_\omega=0.01$.
Note that control is robust to choosing larger bounds on the 
maximal control parameter $\a\ind{max}=0.05$ facilitating fast 
control and leading to faster convergence to the target 
position.
}
\end{figure}

Remarkably, control remains applicable for inhomogeneous rings 
of oscillators with distributed frequencies~$\omega_k$. Note 
that the control perturbations~\eqref{eq:ControlDynamicsGen} are
calculated from the averaged quantity~$\aOPh$. Thus, small 
fluctuations induced by inhomogeneities average out. The resulting
controlled chimera follows the imposed target position even for 
comparatively large standard deviations of the frequency 
distribution; cf.~Figure~\ref{fig:ControlledChimeraInhom}. The 
qualitative impact of control is the same as in homogeneous 
rings. However, if the maximal control parameter~$\a\ind{max}$
is too small, even a controlled chimera may get ``stuck''
while moving towards the target position.

Larger bounds for the control parameter~$\a$ counteract this 
limitation induced by inhomogeneity. In fact, control is not
only robust to choosing $\a\ind{max}>0.015$ but a sufficiently
large value of~$\a\ind{max}$ allows a chimera to be placed 
at an arbitrary position along any inhomogeneous ring. Moreover,
the chimera attains its target position quickly. Carrying out the 
same statistics as previously (i.e., as for assessing the control 
of pseudo random fluctuations for homogeneous rings) reveals
that for sufficiently large control parameters the chimera will
stay on arbitrary targets (not shown). Hence, control renders 
the spatial position of a chimera usable in both homogeneous 
and inhomogeneous systems.


\section{Functional Chimera States}%

\begin{figure}
\begin{center}
\subfloat[Location Encodes Information]{$\begin{array}{c}\includegraphics[scale=\imagescaling]{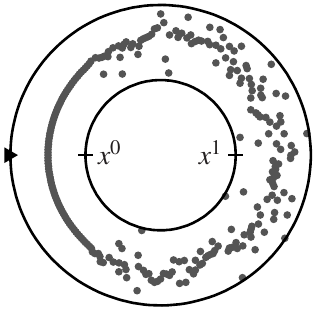}\end{array} := 0\hspace{1.1cm}
\begin{array}{c}\includegraphics[scale=\imagescaling]{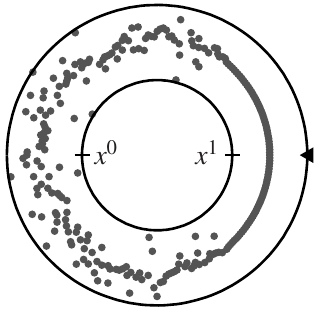}\end{array} := 1$}\\
\medskip
\subfloat[Assignment ``$\B=\A$'' (synchronization)]{$\begin{array}{c}\includegraphics[scale=\imagescaling]{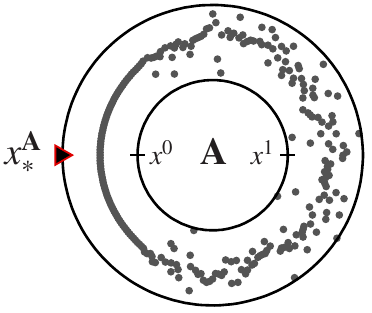}\end{array}\begin{array}{c}\stackrel[\text{coupling}]{\xt^{\B}\left(\vphi^{\A}(t)\right)}{\includegraphics[scale=0.31]{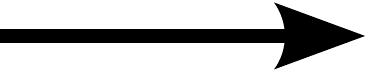}}\end{array}
\begin{array}{c}\includegraphics[scale=\imagescaling]{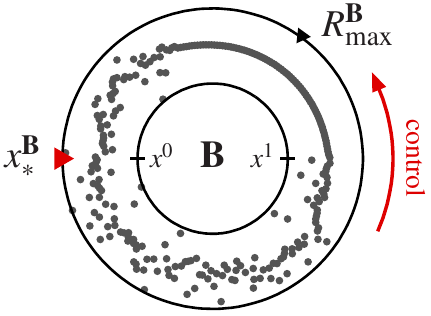}\end{array}$}\\
\medskip
\subfloat[NOT operation ``$\B=\lnot \A$'' (inversion)]{$\begin{array}{c}\includegraphics[scale=\imagescaling]{img/gChimeraCircleLblSyncA}\end{array}\begin{array}{c}\stackrel[\text{coupling}]{\xt^{\B}\left(\vphi^{\A}(t)\right)}{\includegraphics[scale=0.31]{img/sArrow}}\end{array}
\begin{array}{c}\includegraphics[scale=\imagescaling]{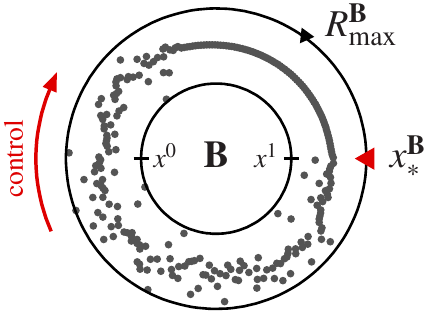}\end{array}$}
\end{center}
\caption{\label{fig:ChimeraComputer}In a digital chimera computer, control 
allows for the spatial position of a chimera to encode information. A chimera 
located at antipodal points $x^0, x^1$ can for example encode bits (Panel~(a)). 
Here the angle denotes the spatial coordinate on the ring~$\S$ and the radius 
the current phase; one obtains the torus by identifying the two boundary 
components of the annulus. By coupling multiple rings through control (black 
arrows), one 
can now realize computations in a chimera computer. 
The current position of a chimera is given by 
$\aOP\ind{max}^X = \argmax_{x\in\S}\aOPh\big(x, \vphi^{X}(t)\big)$, 
$X\in\sset{{\A}, {\B}}$ (black triangles) and coupling between the rings 
(black arrows) is achieved by the dependence of the control target 
(red triangles) $\xt^{\B}$ of Ring~$\B$
on the position $\aOP\ind{max}^{\A}$ of the chimera on Ring~${\A}$, 
cf.~\eqref{eq:DynamicTarget}.
Synchronization of position corresponds to copying bits (Panel~(b)), 
inversion of the position to a NOT gate (Panel~(c)).}
\end{figure}

Control is essential to give chimera states persistent functional meaning. 
Chimera states arise in real physical systems that are related to various 
technological applications. These include collections of mechanical, 
\mbox{(electro-)}chemical, and optical 
systems~\cite{Martens2013, Wickramasinghe2013, Tinsley2012, Hagerstrom2012}.
Chimera control now allows to use the localized nature of a chimera 
state for arbitrary novel applications in these contexts.
As simple example for a technological application of chimera states, 
one may envision a digital chimera computer where spatial location 
directly encodes information. Note that as long as the number of oscillators
is large enough one is not limited to a digital computer with just two states 
but one could also consider an arbitrary number of states up to approximately 
encoding a continuous variable. Take two antipodal points~$x^0, x^1\in\S$ on 
the ring and say that the system is in state~0 
if a chimera is centered at~$x^0$ and in state~1 if it is centered 
at~$x^1$; cf~Figure~\ref{fig:ChimeraComputer}(a). Thus, in this 
setup, the spatial position of a chimera encodes information. With
active control this spatial encoding is reliable because there are 
no random flips between states~0 and~1. Note that only few oscillators 
are necessary to encode information because control reduces the 
pseudo-random fluctuations even in low-dimensional systems.

Control also allows to change the value of the ``bit'' dynamically 
to perform computations. If we take two rings, Ring~$\A$ and Ring~$\B$,
and use the maximum of the order parameter of Ring~$\A$ (with phases 
given by $\vphi^{\A}$) as the target position~$\xt^{\B}$ for Ring~$\B$,
the position of the chimera synchronizes. More explicitly, 
\begin{equation}\label{eq:DynamicTarget}
\xt^{\B}(t) = \argmax_{x\in\S}\aOPh\big(x, \vphi^{\A}(t)\big),
\end{equation}
is the target position for Ring~$\B$ with dynamics given by~\eqref{eq:RingDynamics} 
with coupling kernel~\eqref{eq:CouplingKernel} and 
control~\eqref{eq:DiscrControlScheme}. In terms of the chimera computer, 
this corresponds to an assignment ``$\B=\A$'' or memory copy operation; 
cf.~Figure~\ref{fig:ChimeraComputer}(b). With the minimum 
of~$\aOP(x, \vphi_t^{\A})$ as the target position, the resulting dynamics 
correspond to a NOT operation; cf.~Figure~\ref{fig:ChimeraComputer}(c). 
By coupling multiple rings, one can construct AND and OR gates in a 
similar manner. Here the dynamic target position~\eqref{eq:DynamicTarget} 
is given by a suitable function that depends on the state $\vphi^{\A}(t)$. 
It would be desirable to have a fast, efficient, and natural way to 
determine this target in particular implementations in the future such 
as using adaptive neural networks as a coincidence detector.

Localized dynamical states are directly related to function in neural 
and other biological networks~\cite{Hubel1959, Fyhn2004, Feinerman2008}. On 
the one hand, localized synchrony is generally regarded to play a role for 
example in memory formation~\cite{Fell2011}. On the other hand, localized 
activity at a particular location have been widely studied in spatially 
continuous neural field models as bump states~\cite{Amari1977, Laing2001}.
Neural field models are related to classical pattern forming 
systems~\cite{Ermentrout1998} and stationary localized solutions have been 
given functional interpretation in these models, such as encoding the 
position of a rat's head which can be modulated by inducing asymmetry in 
the coupling~\cite{Zhang1996, Knierim2012}. Chimera states in coupled 
oscillators relate to function both by local synchrony 
(the chimera's synchronized region) as well as by localized activity 
(rotating oscillators make up the incoherent region of a chimera). 
Chimeras and bump states have also been observed in various systems of 
neural oscillatory units with both continuous 
coupling~\cite{Sakaguchi2006a, Omelchenko2013, Hizanidis2013, Laing2014} 
and pulse coupling~\cite{Compte2000, Wildie2012} and have been associated 
with short term memory~\cite{Wimmer2014}. Despite their apparent 
phenomenological similarities to bump states in classical neural field 
models~\cite{Laing2009}, chimera states in coupled oscillators are
mathematically different. Systems of individual coupled oscillators
show multistability of chimeras and the fully synchronized 
state~\cite{Kuramoto2002, Compte2000} and the oscillators rotate rigidly.
Thus, field equations directly derived from collections of oscillators 
contain phase information~\cite{Laing2014} which is crucial to describe 
synchronization. On the other hand, activity described in neural field 
models with just a single variable does not contain any phase information 
whereas the coupling in systems exhibiting chimeras has a phase 
synchronizing effect.

If chimeras as localized states are a feature of biological 
networks, e.g.~\cite{Compte2000, Wimmer2014}, then control is one possible 
mechanism how information is robustly processed in 
these systems. Chimera control allows to both modulate the spatial position of 
a chimera state in finite dimensional systems and keep it as a specified 
location. In contrast to simple information encoding in spatially continuous 
rings~\cite{Zhang1996} with nonautonomous modulation, chimera control---as 
noninvasive feedback control---is a closed loop system where any 
target position can be attained even when 
external input is not constantly available, structural constraints limit 
the maximal asymmetry of the coupling, or the system is incapable of fully 
integrating the input. The control scheme naturally acts as an error corrector 
which counteracts the diffusion of localized patterns in ensembles of finitely 
many units~\cite{Compte2000, Omel'chenko2010} thereby preventing information 
loss. 
Consequently, if even small networks with control exhibit the same structural 
robustness needed for computation in biological systems~\cite{Feinerman2008}
as large networks with high redundancy~\cite{Neumann1956, Compte2000}, we 
may expect to find some form of control in real biological systems.


\section{Discussion}%
Chimera control allows the dynamical modulation of the spatial position
of a chimera state in real time. Control is possible despite the 
multistability with the fully synchronized state, even in small 
finite-dimensional rings with strong low number fluctuations. In contrast 
to other recent applications
of control to chimeras~\cite{Sieber2014}, controlling the chimera as a 
whole is the first step towards making use of chimera dynamics in 
practical applications as illustrated by the simple chimera computer. 
Apart from applications, control is relevant for implementation in experimental 
setups. On the one hand, control can directly be applied to a number of 
the current experimental realizations of chimera states such 
as~\cite{Tinsley2012, Hagerstrom2012}. 
In these setups, implementation is straightforward since the coupling is 
computer mediated. On the other hand, control remains applicable  
in more general experimental contexts beyond computer mediated coupling.
Oscillators may be coupled by immersing them in a common reactive-diffusive 
medium~\cite{Shima2004}. Subjecting the medium to a advective concentration 
gradient (due to a sink or source) may give rise to an exponential coupling 
kernel~\eqref{eq:CouplingKernel}: when the time scale characteristic to the 
medium is rapid compared to that of the oscillators, an adiabatic solution is 
viable yielding the asymmetric coupling~\eqref{eq:CouplingKernel},
see~\cite{Shima2004,Bick2014,Siebert2014}.
Since a nonzero advective gradient yields an asymmetric coupling, control can 
be realized by modulating the strength of the gradient. Setups with a common 
medium have been studied in synthetic biology where oscillating cells 
communicate via quorum-sensing~\cite{Danino2010} and can be subjected 
to an advective currents~\cite{Lang2011}. Similar systems could be implemented 
using yeast cells under glycolysis~\cite{Dano1999,Weber2012}, or diffusively 
coupled chemical oscillators in microfluidic assemblies~\cite{Toiya2008,Thutupalli2013}. 
Hence, we anticipate our control strategy to also find direct application in both 
technological and biological experimental setups.
Control may also play an important role in natural biological settings
as already discussed in the section above.

Remarkably, chimera control is robust with respect to perturbations 
of the system. Chimera states persist in non-locally coupled rings of 
nonidentical oscillators~\cite{Laing2009c, Laing2012} and can be 
controlled; cf.~Figure~\ref{fig:ControlledChimeraInhom}. In fact, 
chimera control acts in two ways. If the oscillators are (almost)
identical, then control suppresses the finite size fluctuations. 
Increasing inhomogeneity reduces fluctuations but also restricts 
uncontrolled chimeras to stable locations with respect to movement 
along the ring~$\S$. Control eliminates this limitation for inhomogeneous 
rings and allows chimeras to be placed at any position. This indicates
that chimera control remains applicable in more general oscillator 
models, for example to suppress drift~\cite{Compte2000}. Note that our control is noninvasive in the sense that the control 
signal vanishes on average upon attaining the target position; 
cf.~Equation~(\ref{fig:ControlledChimera}). As a 
result, chimera control is also robust with respect to larger values 
of the symmetry parameter~$\a$ yielding chimeras which attain 
their target position very quickly as indicated in 
Figure~\ref{fig:ControlledChimeraInhom}.

The gradient based control approach immediately extends to higher 
dimensional systems. The only requirement for a successful 
implementation is the availability of an accessible control 
parameter which induces drift. Preliminary numerical simulations 
indicate spiral wave chimeras~\cite{Shima2004, Martens2010}, spiral 
waves with an incoherent core, may exhibit spatial drift. 
Thus, an implementation of control for two dimensional chimera 
states is within direct reach. Gradient dynamics are a relatively naive 
control approach; here it serves as a proof of principle. Given that
there the asymmetry is an accessible control parameter and the
local order parameter an objective function, one would eventually 
like to see more sophisticated control schemes implemented, for 
example speed gradient control~\cite{Fradkov1998}.

In summary, chimera control is a robust control scheme to control
the spatial position of a chimera state and reliably maintain its 
position even for small numbers of oscillators that may be 
nonidentical. Note that chimera control is not limited to the control 
of the position of the synchronized region of a chimera. The control 
scheme presented here may be 
applied if there is a relationship between a control parameter and 
$Q$-traveling solutions for a suitable observable~$Q$. Developing 
novel applications based on controlled chimeras, applying the presented 
control scheme to experimental setups, and studying its relevance in 
biological settings 
provide exciting 
directions for future research.

\section*{Acknowledgements}%

The authors would like to thank M. Field, C. Laing, Yu. Maistrenko, and M. Timme for helpful discussions. Moreover, the authors would like to thank all anonymous referees for helping to improve the presentation of our results and pointing out further references relevant for this work. CB acknowledges support by NSF grant DMS--1265253 and partially by BMBF grant 01GQ1005B. The research leading to these results has received funding from the People Programme (Marie Curie Actions) of the European Union's Seventh Framework Programme (FP7/2007-2013) under REA grant agreement n\textsuperscript{o} 626111 (CB). The work is part of the Dynamical Systems Interdisciplinary Network, University of Copenhagen (EAM).


\section*{References}


\bibliographystyle{unsrt}
\def\urlprefix{}
\def\url#1{}

\bibliography{citations} 

\end{document}